\documentclass[a4paper,11pt]{article}
\usepackage{graphicx}
\usepackage{graphics}
\title{Is there  model-independent evidence of the two-photon-exchange effect in the
  electron-proton elastic scattering cross section?}

\author{\bf Yu Chun Chen$^1$,
Chung Wen Kao$^2$, and Shin Nan Yang$^1$ \\
$^1$Department of Physics and Center for Theoretical Sciences,
National Taiwan University,\\
Taipei 10617, Taiwan,\\
$^2$Department of Physics, Chung-Yuan Christian University,
Chung-Li 320, Taiwan\\
}
\begin{document}
\maketitle
\begin{abstract}
We re-analyze the data of the elastic electron proton scattering
to look for model-independent evidence of the two-photon-exchange
(TPE) effect. In contrast to previous analyses, TPE effect is
parametrized in forms which are free of kinematical-singularity,
in addition to being consistent with the constraint derived from
crossing symmetry and the charge conjugation. Moreover, we fix the
value of $R=G_E/G_M$ as determined from the data of the
polarization transfer experiment. We find that, at high $Q^2 \geq
2 \, GeV^2$ values, the contribution of the TPE effect to the
slope of $\sigma_R$   vs. $\varepsilon$ is large and comparable
with that arising from $G_{E}$. It also behaves quasi-linearly in
the region of current data, namely, in the range of $0.2 <
\varepsilon < 0.95$. Hence the fact that the current elastic $ep$
cross section data shows little nonlinearity with respect to
$\varepsilon$ can not be used to exclude the presence of the TPE
effect. More precise data at extreme angles will be crucial for a
model-independent extraction of the TPE effect.\\

\end{abstract}
\par
\newpage

The electric and magnetic form factors $G_E$ and $G_M$ of the
nucleon contain essential piece of our knowledge of the nucleon
structure. Traditionally, Rosenbluth method is used to extract $G_E$
and $G_M$ of the proton from the differential cross section of the
electron-proton ($ep$) scattering. Within the one-photon-exchange
(OPE) approximation, one has
\begin{equation}
\sigma_{R}(Q^2,\varepsilon)\equiv
\frac{d\sigma}{d\Omega_{Lab}}\frac{\varepsilon(1+\tau)}{\tau\sigma_{Mott}}
=G_M^{2}(Q^2)+\frac{\varepsilon}{\tau}G_E^{2}(Q^2), \label{eq:sred}
\end{equation}
where $\tau=Q^2/4M_{p}^{2}$ and $1/\varepsilon \equiv
1+2(1+\tau)\tan^2\theta_{Lab}/2$, with $\theta_{Lab}$ the laboratory
scattering angle such that $0\leq \varepsilon \leq1$.
$\sigma_{Mott}$ is the Mott cross section. At a fixed value of
$Q^2$, the slope and the intercept of $\sigma_{R}(\varepsilon) \,
vs. \,\varepsilon$ is directly related to $G_{E}$ and $G_{M}$,
respectively. Through this method it is determined that
\begin{equation}
\mu_{p}R=\mu_{p} G_{E}(Q^2)/G_{M}(Q^2)\approx 1,
\label{eq:Rosenbluth}
\end{equation}
where $\mu_{p}=2.79$ is the proton anomalous magnetic moment.

Another way to extract the ratio $R = G_E/G_M$ is to measure the
polarizations of the recoil proton in the elastic $ep$ scattering.
Within the OPE approximation, the values of $R$ can be extracted
according to the following formula \cite{Akhiezer74}:
\begin{equation}
R=-\sqrt{\frac{\tau(1+\varepsilon)}{2\varepsilon}}\cdot
\frac{P_t}{P_l}, \label{eq:LT}
\end{equation}
where $P_l$ and $P_t$ are the polarizations, parallel and
perpendicular to the proton momentum in the scattering plane,
respectively. This method has several advantages over the
traditional Rosenbluth separation method. The reason is that the
ratio of the two simultaneously measured polarization components
greatly reduces systematic uncertainties. For instance, a detailed
knowledge of the spectrometer acceptance which plagues the cross
section measurements is not required here. Furthermore, the detailed
knowledge of the beam polarization and the the polarimeter analyzing
power are also not needed because both  quantities cancel in
measuring the ratio of the form factors. Therefore this method has
been considered to be more reliable.

Recent polarization transfer experiments at Jefferson Lab
\cite{Jones00,Gayou02} give, however,
\begin{equation}
\mu_p R=\mu_p
G_{E}(Q^2)/G_{M}(Q^2)=1-0.13(Q^2~[\mbox{GeV}^2]-0.04),
\label{eq:brash}
\end{equation}
which differs substantially  from the Rosenbluth results of Eq.
(\ref{eq:Rosenbluth}) at high $Q^2$.  It was suspected that there
is a systematic uncertainty common to all Rosenbluth data.
However, recent measurements of high $Q^2$ $ep$ elastic cross
section at the Jefferson Laboratory (JLab)
\cite{Christy04,Qattan05} confirmed the earlier Rosenbluth
measurements from SLAC \cite{Walker94,Andivahis94} in the same
$Q^2$ region.

It has been proposed \cite{Guichon03,Blunden03,Chen04} that this
mismatch can be explained by the two-photon-exchange (TPE) effect
not fully accounted for in the standard radiative corrections
procedure of Mo and Tsai \cite{MoTsai}. So far the only
observation of possible TPE effect is in the asymmetry of
scattering of transversely polarized electrons from protons. This
observable should vanish within the OPE approximation but both
SAMPLE experiment at Bates \cite{SAMPLE} and PVA4 Collaboration at
Mainz \cite{Maas05} have observed small but non-zero results. This
observable is related to the imaginary part of two-photon exchange
amplitude. On the other hand, the unpolarized cross section
depends only on the real part of the two-photon exchange amplitude
which does not, in general, vanish. It is hence reasonable to
expect that the TPE effect would contribute to $\sigma_{R}$. The
crucial question is whether this effect is large enough to explain
the discrepancy between the results of the Rosenbluth and the
polarization transfer methods for the ratio $R$.

To be more specific, let us take a closer look on how the TPE
effect would modify the Rosenbluth formula. Including the TPE
effect, Eq.~(\ref{eq:sred}) is rewritten in the following general
form \cite{Guichon03}:
\begin{equation}
\sigma_{R}(Q^2, \varepsilon)= G_M^{2}(Q^2)+
\frac{\varepsilon}{\tau}G_E^{2}(Q^2) + F(Q^2,\varepsilon),
\label{eq:sred2}
\end{equation}
where $F(Q^2,\varepsilon)$ is a real function describing the
effect of the $1\gamma\otimes 2\gamma$ interference. Although the
magnitude of $F$ is suppressed by $\alpha_{em}=1/137$ compared to
the OPE effect, the $\varepsilon$ dependence of
$F(Q^2,\varepsilon)$ is not necessarily small in comparison with
$\varepsilon G_E^{2}/\tau$, particularly at higher $Q^2$ region
where $G_{E}$ and $1/\tau$ are both small. Hence the slope of the
$\sigma_{R}(\varepsilon)$ is no longer just $G_E^{2}/\tau$ but
also includes the linear part of the TPE effect. As a result, the
validity of Eq.~(\ref{eq:sred}) becomes questionable and the
extraction of $G_E$ and $G_M$ from $\sigma_{R}(\varepsilon)$ needs
more sophisticated consideration and the values of $R=G_{E}/G_{M}$
obtained via the Rosenbluth method should also be modified.

There are currently two ways to study  the TPE effect in
$\sigma_{R}$. The first one is to estimate $F(Q^2,\varepsilon)$ by
calculating the two-photon-exchange $ep$ scattering amplitude with
some simple hadronic or partonic models
\cite{Blunden03,Chen04,Blunden05}. However, so far there is no
complete calculation which is valid at all kinematics. The second
way is to analyze the experimental data to search for the
nonlinearity in $\sigma_{R}$ with respect to $\varepsilon$ which
is regarded as the evidence of the TPE effect~\cite{Tg05,Tv06}
since $F(Q^2,\varepsilon)$ is, in general, not a linear function
of $\varepsilon$.

All current model calculations \cite{Blunden03,Chen04,Blunden05}
show that the existing discrepancy between the Rosenbluth and the
polarization transfer methods can at least be partially explained
by the TPE effect~\cite{Guichon03,Chen04}. However, several
groups~\cite{Tg05,Tv06} have claimed that they cannot find any
conclusive evidence for the the nonlinearity of
$\sigma_R(\varepsilon)$ in their data analyses. The analysis of
Ref.~\cite{Tv06} set tight limits on the size of the nonlinearity
of $\sigma_{R}(\varepsilon)$. Nevertheless, these limits are still
consistent with the predictions of the model calculations. In
fact, this issue was already discussed in \cite{Guichon03}. It was
speculated that there may be a simple explanation of the fact the
Rosenbluth plot looks linear even if it is strongly affected by
the TPE effect. Namely, even if the TPE effect in the cross
section is not a linear function of $\varepsilon$, it is still
possible for $F$ to behave quasi-linearly within the present
experimental uncertainty in the current data region ($0.2 <
\varepsilon < 0.95$). If this is the case, both of the success of
Rosenbluth separation in the past and the newly discovered
discrepancy can be explained because the $\sigma_{R}(\varepsilon)$
in the certain range of $\varepsilon$ could look like a straight
line with its slope a sum of the OPE and TPE effects.

To explore this possibility, one needs to look for some appropriate
parametrization of the TPE effect to account for the data of the
cross section of the elastic $ep$ scattering. Such a parametrization
has to possess general characteristics of the TPE mechanism, as
derived from charge conjugation and crossing symmetry in Ref.
\cite{Rekalo04a}. Namely, the last term $F$ in  Eq. (\ref{eq:sred2})
which corresponds to the $1\gamma\otimes 2\gamma$ interference
effects should, at a fixed value of $Q^2$, satisfy the following
constraint
\begin{equation}
F(Q^2,y) = - F(Q^2,-y), \label{eq:con}
\end{equation}
where $y=\sqrt{\displaystyle\frac{1-\varepsilon}{1+\varepsilon}}$.
In \cite{Tg05}, $F(Q^2,\varepsilon)$ is assumed to take the form
\begin{equation}
F(Q^2,\varepsilon) \sim  \frac{\varepsilon}{y} \mbox{ or }
\frac{1}{y}. \label{eq:sfit}
\end{equation}
However, it suffers from the problem that $1/y$ diverges when $y$
approaches zero, i.e., $\theta_{Lab}\rightarrow 0$, and is hence
not acceptable.

In this letter, we study the TPE effect in
$\sigma_{R}(\varepsilon)$ directly through data analysis. We
choose the data from \cite{Andivahis94} in order to avoid
complication due to  different normalizations of different
experiments. According to the general argument of
~\cite{Guichon03}, $R= G_E/G_M$ determined in the polarization
transfer method is little affected by the TPE effect, at least in
the range of $Q^2$ which has been determined till now. As a
result, it is important to fix the values of $R$ as obtained from
polarization transfer experiment in any analysis. With the use of
$R$ determined from  polarization transfer experiments, Eq.
(\ref{eq:sred2}) can be expressed as:
\begin{equation}
\sigma_{R} = G_M^{2}(Q^2)\left(1 + \frac{\varepsilon}{\tau}
R^2\right) +F(Q^2,\varepsilon). \label{eq:sred3}
\end{equation}
In other words, in Eq. (\ref{eq:sred3}) the input parameters are now
$G_M$ and $F$ with $R$ given by Eq. (\ref{eq:brash}).

\begin{table}[h]
\begin{tabular}{|c|c|c|c|c|c|}
\hline $Q^2$[GeV$^2$] & $\displaystyle G_M^{2}$($10^{-3}$)&
$\frac{1}{\tau}G_E^{2}(10^{-3})$& $ A(10^{-3})$
& $\chi^2 $ & $N_{ points}$\\
\hline
1.75& 62.96 & 9.825 &   -3.040  &  0.194 &4 \\
2.50 &21.58 & 1.803 & -0.840  & 0.110  &7 \\
3.25&9.340 & 0.441& -0.550 & 0.115 &5 \\
4.00 & 4.578 & 0.122 & -0.280 &0.313&6 \\
5.00 & 2.052 & 0.023& -0.130 &0.637&5 \\
\hline
\end{tabular}
\caption{The values of the form factors and  coefficient A obtained
in the fit II, Eq. (\ref{eq:sred4}), with $B(Q^2)$ assumed to be
$0$.} \label{tab:B=0}
\end{table}

\begin{table}[h]
\begin{tabular}{|c|c|c|c|c|c|}
\hline $Q^2$[GeV$^2$] & $\displaystyle G_M^{2}$($10^{-3}$) &
$\frac{1}{\tau}G_E^{2}$ ($10^{-3}$) & $B(10^{-3})$
& $\chi^2 $ & $N_{ points}$ \\
\hline
1.75& 62.38& 9.734& -3.950  & 0.060 &4 \\
2.50 &21.34& 1.783 & -0.890 & 0.141  &7 \\
3.25&9.176& 0.433& -0.580 & 0.089 &5 \\
4.00 & 4.499& 0.120 & -0.290 &0.433&6\\
5.00 & 2.016& 0.023& -0.140 &0.556&5 \\
\hline
\end{tabular}
\caption{The values of the form factors and  coefficient B obtained
in the fit II, Eq. (\ref{eq:sred4}), with $A(Q^2)$ assumed to be
$0$.} \label{tab:A=0}
\end{table}
\begin{table}[h]
\begin{tabular}{|c|c|c|c|c|}
\hline $Q^2$[GeV$^2$] & $\displaystyle G_M^{2}$($10^{-3}$)
& $\frac{1}{\tau}G_E^{2}$ ($10^{-3}$)  & $\chi^2$ & $ N_{points}$ \\
\hline
1.75& 59.90 & 12.52 & 0.147&4\\
2.50 &20.76 & 2.493 &0.103&7\\
3.25&8.795& 0.897&0.097&5\\
4.00 & 4.298& 0.359 & 0.340&6 \\
5.00 & 1.925& 0.128&0.644&5 \\
\hline
\end{tabular}
\caption{The values of the form factors obtained with the Rosenbluth
fit (fit I). The $\chi^2$ values are also presented.} \label{tab:0}
\end{table}
In contrast to Eq. (\ref{eq:sfit}) which is singular at $y=0$
($\epsilon =1)$, we will assume $F(Q^2,y)$ is analytic around $y=0$.
The most natural choice of the parametrization of $F(y)$ will then
be a polynomial of $y$ as would be obtained in a Taylor expansion.
In addition, $y^{n}, n\geq 1$ increases when $\varepsilon$
decreases, vanishes at $\varepsilon=1\, (y=0)$ and stays finite when
$\varepsilon=0 \,(y=1)$. This feature agrees with the results of the
model calculations \cite{Blunden03,Chen04}. Therefore one may
parameterize Eq. (\ref{eq:sred3}) as a function of three parameters,
$G_M$, $A$, and $B$:
\begin{equation}
\sigma_{R} = G_M^{2}(Q^2)\left(1 + \frac{\varepsilon}{\tau} R^2
\right) + A(Q^2) y + B(Q^2) y^3. \label{eq:sred4}
\end{equation}

\begin{table}
\begin{tabular}{|c|c|c|c|c|c|c|c|}
\hline $Q^2$[GeV$^2$] & $G_M^{2}(10^{-3})$
& $\frac{1}{\tau}G_E^{2} (10^{-3})$ &  $A(10^{-3})$ & $B(10^{-3})$ & $P_2(\%)$&$\chi^2 $ & $ N_{points}$ \\
\hline
1.75& 62.68 & 9.75 &  $-1.533$ & -1.943 & -3.15& 0.110 &4 \\
2.50 &21.55 & 1.80 & -0.529& $-0.670$ & -3.28&0.177&7 \\
3.25&9.236& 0.436& -0.228& -0.289 & -3.36& 0.095 &5 \\
4.00 & 4.526& 0.120 & -0.114& $-0.145$ & -3.44& 0.366 &6\\
5.00 & 2.027& 0.023& -0.053& -0.068 &-3.69&0.582 &5 \\
\hline
\end{tabular}
\caption{The values of the form factors and  coefficients A and B
obtained in the fit II, Eq. (\ref{eq:sred4}). The $\chi^2$ values of
this fits are also presented.} \label{tab:fitB}
\end{table}

\begin{table}
\begin{tabular}{|c|c|c|c|c|c|c|c|}
\hline $Q^2$[GeV$^2$] & $\displaystyle G_M^{2}$ ($10^{-3}$)
& $\frac{1}{\tau}G_E^{2}(10^{-3})$ & $\hat{A}(10^{-3})$ & $\hat{B}$($10^{-3}$) &$ P_2(\%)$&$\chi^2$ & $N_{ points}$\\
\hline
1.75& 63.88& 9.982 &   -4.261  & -1.422& -0.99&0.191  &4 \\
2.50 &22.00 & 1.841 & -1.470  & -0.491 & -1.03&0.162 &7 \\
3.25&9.436& 0.446& -0.635 &-0.212 & -1.06& 0.122 &5 \\
4.00& 4.625 & 0.123 & -0.317 &-0.106 &-1.09& 0.329&6 \\
5.00 & 2.073 & 0.024& -0.147 &-0.049& -1.14&0.629&5 \\
\hline
\end{tabular}
\caption{The values of the form factors and coefficients $\hat A$
and $\hat B$ obtained within the fit III, Eq. (\ref{eq:sred6}). The
$\chi^2$ values of this fits are also presented.} \label{tab:fitC}
\end{table}

It is natural to expect that both $A(Q^2)$ and $B(Q^2)$ are smooth
functions of $Q^2$. Tables \ref{tab:B=0} and \ref{tab:A=0} show the
results of the fits where either $A$ or $B$ is set to be zero. The
$\chi^2$'s obtained with these parametrizations are as small as
those given in the Rosenbluth fit of Table \ref{tab:0}. The values
of $A$ or $B$ decreases but the ratio between $A$ (or $B$) and
$G_{E}^{2}/\tau$ increases when $Q^2$ increases. It is difficult to
judge which one gives a better fit because at $Q^2$=2.5 and 4.0
GeV$^{2}$ the $\chi^2$  with $B=0$ is smaller, while for $Q^2$ =
1.75, 3.25, and 5.0 GeV$^2$ the fit with $A=0$ gives smaller
$\chi^2$. Therefore it is reasonable to keep both $A$ and $B$ in
Eq.~(\ref{eq:sred4}). If we further parametrize $A(Q^2)$ and
$B(Q^2)$ by
\begin{equation}
A(Q^2)=\alpha G_{D}^{2}(Q^2),\,\,\,B(Q^2)=\beta
G_{D}^{2}(Q^2),\,\,\, G_{D}=\frac{1}{(1+Q^2/0.71)^{2}}, \label{AB}
\end{equation}
then the number of parameters reduces from fifteen to seven (the
values of $G_{M}$ at five different $Q^2$ values and two constants
$\alpha, \beta$). The result of such a fit is presented in Table
\ref{tab:fitB} where $\alpha=-0.221$ and $\beta=-0.28$.

It is also possible to  parameterize $F$ by functions which are
not analytic at $y=0$ but behave smoothly in the region of $0\leq
y \leq 1$ ($0 \leq \varepsilon \leq 1$). The logarithmic or
double-logarithmic functions which are not analytic at $y=0$, are
such types of functions often appear in the loop diagrams
\cite{Chen04}. So $y \ln|y|$ or $y (\ln|y|)^2$ should also be
possible choices for the TPE contribution. However, the logarithm
type functions will be zero when $y$ approaches one. This feature
does not agree with the model calculations \cite{Chen04}.
Therefore we assume that $F$ should be parameterized by some
combinations of a polynomial of $y$ and some logarithm type
functions of $y$ such as:
\begin{equation}
\sigma_{R} = G_M^{2}(Q^2)\left(1 + \frac{\varepsilon}{\tau} R^2
\right) + \hat{A}(Q^2) y +\hat{B}(Q^2) y (\ln |y|)^2.
\label{eq:sred6}
\end{equation}
If we again parametrize $\hat{A}(Q^2)$ and $\hat{B}(Q^2)$ by
\begin{equation}
\hat{A}(Q^2)=\hat{\alpha}G_{D}^{2}(Q^2),\,\,\,\hat{B}(Q^2)=\hat{\beta}G_{D}^{2}(Q^2),\,\,\,
G_{D}=\frac{1}{(1+Q^2/0.71)^{2}},\label{AC}
\end{equation}
we obtain results as given in Table \ref{tab:fitC} where
$\hat{\alpha}=-0.614$ and $\hat{\beta}=-0.205.$

Let us now compare the results of three different fits, the
Rosenbluth fits of Eq. (\ref{eq:sred}), with $G_{M}$ and $G_{E}$ as
parameters (fit I), polynomial fit of Eq. (\ref{eq:sred4}) with
parameters $G_{M}$ and $A$ and $B$ (fit II), and the polynomial plus
logarithmic function fit of Eq. (\ref{eq:sred6}) with $G_{M}$, $\hat
A$, and $\hat B$ (fit III) as the parameters. Their results are
presented in Tables \ref{tab:0},  \ref{tab:fitB}, and
\ref{tab:fitC}, respectively. In fit II,  $A$ and $B$ are
parametrized according to Eq. (\ref{AB}) and in fit III, $\hat{A}$
and $\hat{B}$ are given by Eq. (\ref{AC}). It can be seen that the
resultant values of $\chi^2$ in fits II and III are as small as or
even smaller than the ones of the Rosenbluth fit. Hence, our choices
of the parametrization of the TPE effect $F$ indeed can explain the
$ep$ cross section data with the values of $R=G_E/G_M$ fixed by the
polarization transfer experiment data. It is worth emphasizing that
there are ten input parameters in the fit I but only seven input
parameters in the fit II and III.

Even though the ansatz used in the fits II and III are quite
different, there are some common features. First, the effect of the
$1\gamma\otimes 2\gamma$ interference is destructive, namely,
contribution of $F$ is always negative. Consequently, the values of
$G_{M}$ obtained in fits II and III are always larger than those of
the Rosenbluth fit. In fits II and III, the values of $G_{M}$ are
enhanced by about $2\%$ which is very small but nevertheless
comparable with the experimental uncertainty. In general the values
of $G_{M}$ in the fit III are larger than the ones in the fit II. In
addition, the values of $G_{E}$ at higher $Q^2 \geq 2 \,GeV^{2}$ are
much smaller than the results of the Rosenbluth fit because the
slope of $\sigma_{R}(\varepsilon)$ receives large contribution from
the TPE effect, as will be explained in more details later.

\begin{figure}
\centering
\includegraphics[width=10cm,height=16cm]{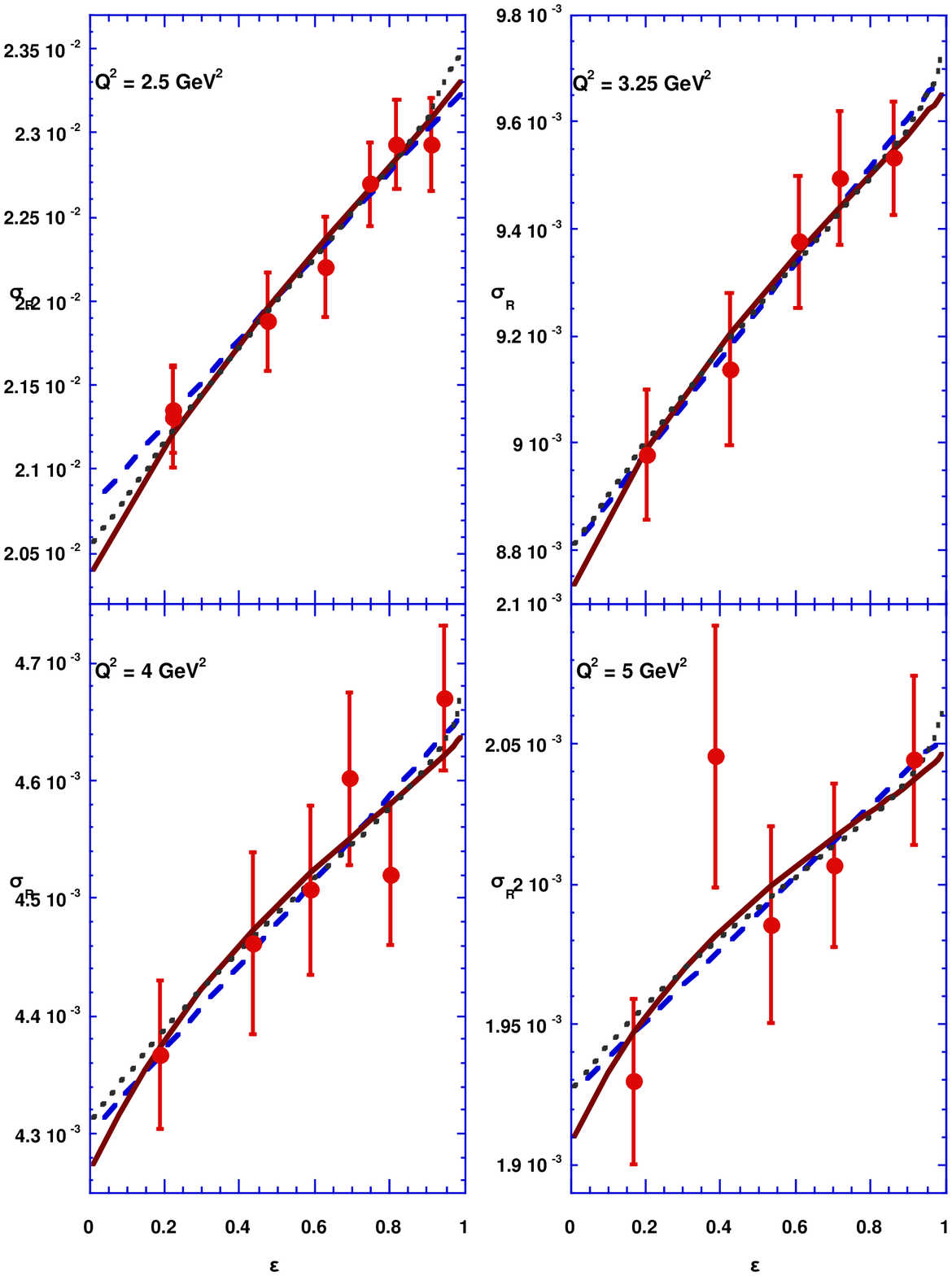}
\caption{The reduced cross section $\sigma_{R}$ at $Q^2=1.75$
$GeV^2$. The dashed line is the Rosenbluth fit. The solid line
represents the fit II. The dotted line represents the fit III.}
\label{Fig:all}
\end{figure}

Fig. \ref{Fig:all}
shows $\sigma_{R}(\varepsilon)$ at $Q^2 =$ 2.5, 3.25, 4.0, and 5.0
GeV$^{2}$, respectively, for the data from Ref.~\cite{Andivahis94}.
The dashed, solid, and dotted lines, correspond to the results of
the fits I, II, and III, respectively. The dashed lines are straight
lines and the dotted lines and the solid lines have small
curvatures. One sees that both of the solid and dotted lines, in the
range of $0.2 < \varepsilon < 0.95$, are close to the dashed lines
and both of them show very little nonlinearity which agrees with the
previous analyses \cite{Tg05,Tv06}. It is interesting to see that
even though the $1\gamma\otimes 2\gamma$ interference term $F$ is in
principle not a linear function of $\varepsilon$,  it contributes
significantly to the slope but very little to the curvature in fits
II and III. Furthermore, we observe that outside the current data
region, the dotted lines show larger nonlinearity than the solid
lines at very high $\varepsilon$ region ($\varepsilon > 0.95)$ .
Hence, if the logarithm type functions play important role then the
nonlinearity of $\sigma_{R}$ will become more pronounced at
$\varepsilon > 0.95$. On the other hand, in the region of
$\varepsilon < 0.2$, the solid lines show larger nonlinearity than
the dotted lines. The nonlinearity of both solid and  dotted lines
are too small as compared to the uncertainty of the current data.

To estimate the contribution of the TPE effect to the slope of
$\sigma_{R}(\varepsilon)$, one can expand Eqs. (\ref{eq:sred4})
and (\ref{eq:sred6}) around $\varepsilon=1/2$ as in ~\cite{Tv06}:

\begin{equation}
\sigma_{R}(\varepsilon) =
\left(G_{M}^{2}+\frac{G_{E}^{2}}{2\tau}+C_{0}\right)
+\left(\frac{G_{E}^{2}}{\tau}+C_{1} \right)
\left(\varepsilon-\frac{1}{2}\right)+C_{2}\left(\varepsilon-\frac{1}{2}\right)^2
+{\cal O}((\varepsilon-\frac{1}{2})^{3}), \label{eq:sred5}
\end{equation}
where $C_{0},C_{1},$ and $C_{2}$ arise from the interference term
$F$. In fit II, one has
\begin{equation}
C_{0}=\frac{A+3B}{\sqrt{3}},\,\,C_{1}=\frac{4(A+B)}{3\sqrt{3}},\,\,\,C_{2}
=\frac{16B}{9\sqrt{3}},
\end{equation}
while in the fit III, one has
\begin{equation}
C_{0}=\frac{4\hat{A}+\ln^{2}3\hat{B}}{4\sqrt{3}},\,\,C_{1}=\frac{4\hat{A}+(4\ln3-\ln^{2}3)\hat{B}}{3\sqrt{3}},\,\,\,C_{2}
=\frac{(16-8\ln3)\hat{B}}{9\sqrt{3}}.
\end{equation}

\begin{figure}
\centering
\includegraphics[width=12cm,height=8.5cm]{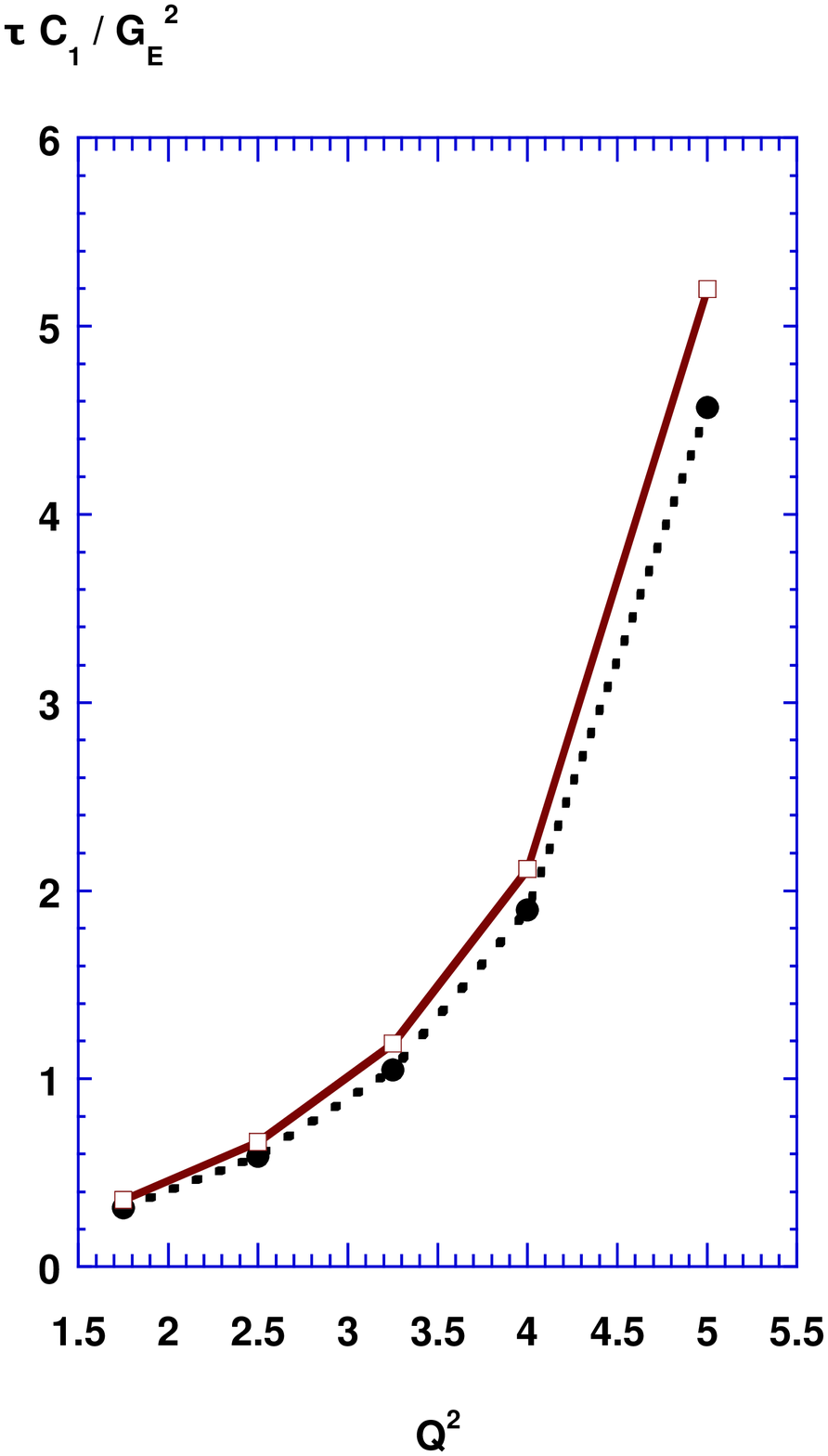}
\caption{The values of $C_{1}/(\frac{1}{\tau}G_{E}^{2})$ in the fit
II and III. The solid line represents the fit II and the dashed line
represents the fit III. } \label{Fig:C1}
\end{figure}

The value of $C_{0}$ is related to the contribution of the TPE
effect to the intercept of $\sigma_{R}$. $C_{1}$ represents the
TPE contribution to the slope of $\sigma_{R}$. $C_{2}$ reflects
the curvature of $\sigma_{R}$. Fig.~\ref{Fig:C1} shows  the value
of $C_{1}/(G_{E}^{2}/\tau)$ as a function of $Q^2$, which
represents the ratio between the contributions of the TPE and the
OPE effects to the slope of $\sigma_R$, for both fits II and III.
One sees that the value of $C_{1}/(G_{E}^{2}/\tau)$ increases from
a value of $\sim 0.3$ at $Q^2=1.75\, GeV^2$ to about $4-6$ at
$Q^2=5\, GeV^2$, a huge effect.  On the other hand, the value of
$C_{0}/G_{M}^{2}$ remains small. We stress that it is $G_{E}$
term, not $G_{M}$ term one should look for the TPE effect because
the contribution of the TPE effect to the cross section is only
few percents but it dramatically changes the slope of the
Rosenbluth plots at high $Q^2$.
If one neglects the TPE effect, the values of $G_{E}$ will be
greatly overestimated at large $Q^2$. It is interesting to observe
that $A \,(\hat A)$ never appears in $C_2$, the coefficient of
$(\varepsilon-1/2)^2$. In \cite{Tv06}, the following form has been
adopted in their analysis,
\begin{equation}
\sigma_{R}=P_{0}\cdot
\left[1+P_{1}(\varepsilon-\frac{1}{2})+P_{2}(\varepsilon-\frac{1}{2})^2\right],
\end{equation}
and they obtained a very small $P_{2}$. It is not difficult to
understand this result because $P_{2}$ corresponds to $C_2/C_0$
which is proportional to the ratio between $B \,(\hat B)$ and
$G_M^2$. In addition, one notes that if $B \,(\hat B)=0$,  then
$P_{2}$ is identically equal to zero. In that case one needs to
expand $\sigma_R$ up to ${\cal O}(\varepsilon^3)$. The values of
$P_2$ are around a few percent as seen in Tables  \ref{tab:fitB} and
\ref{tab:fitC}, which agree with the findings in \cite{Tg05,Tv06}.
In general the values of $P_2$ obtained in fit II are about three
times as large as those of the fit III and they increase slowly with
$Q^2$.

\begin{figure}
\includegraphics[width=8cm,height=8.5cm]{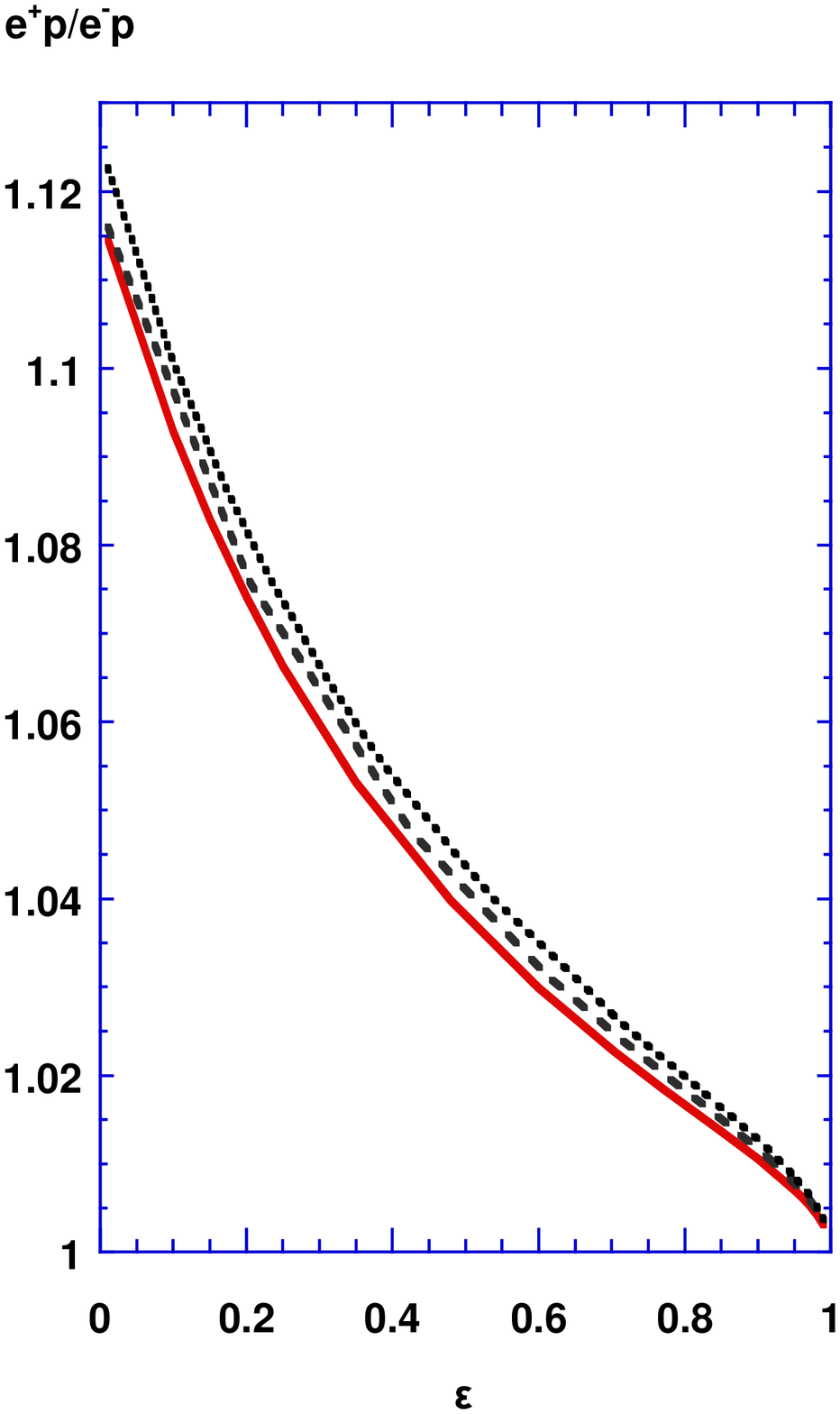}
\includegraphics[width=8cm,height=8.5cm]{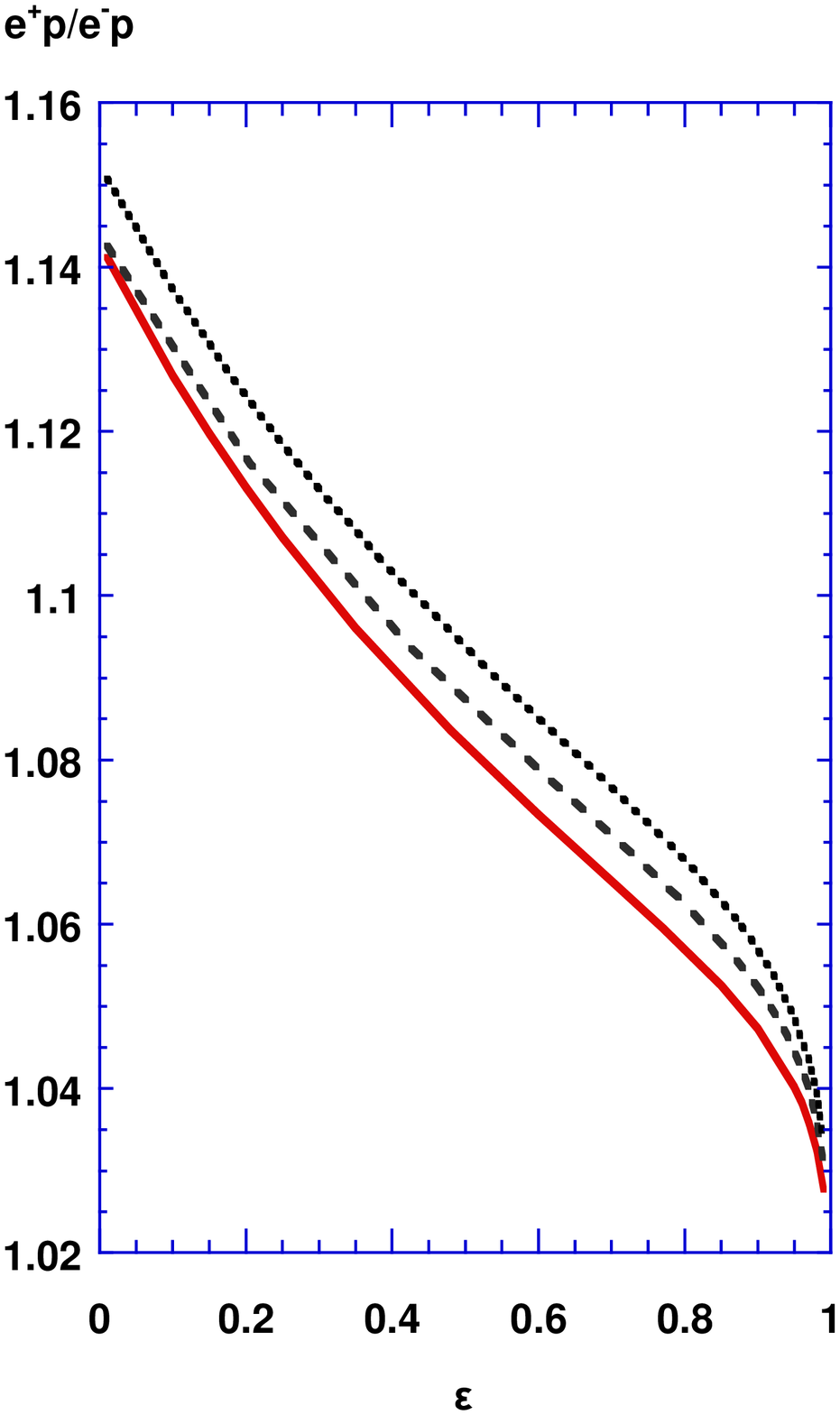}
\caption{The ratio $R^{\pm}$ of the positron-proton and
electron-proton scattering cross section according to the fit II
(left panel) and fit III (right panel). The dotted, dashed, and
solid curves correspond to results of $Q^2 = 1.75, 3.25,$ and $5.0
\, GeV^2$, respectively.} \label{Fig:R}
\end{figure}

For positron-proton scattering the interference term between the
one-photon and the two-photon exchange amplitudes changes sign and
yields a ratio $R^{\pm}=\sigma(e^{+}p)/\sigma(e^{-}p)\approx 1+4
Re(A^{2\gamma}/A^{1\gamma})$, where $A^{2\gamma}$ and
$A^{1\gamma}$ are the two-photon and one-photon exchange
amplitudes, respectively. Owing to the low luminosity of the
secondary positron beams, almost all the comparisons were made at
very low $Q^2$, or at small scattering angels, corresponding to
$\varepsilon>0.7$. New positron measurements \cite{Arrington04}
will be performed in the near future to measure the TPE effect at
larger angels and moderate $Q^2$ values. Here we present our
$R^{\pm}$ in Fig. \ref{Fig:R} for fits II and III, where dotted,
dashed, and solid lines correspond to results of $Q^2 = 1.75,
3.25,$ and $5 \,GeV^2$, respectively. Indeed at low $Q^2$ and high
$\varepsilon$ region, the $R^{\pm}$ is smaller than a few percent.
The values of $R^{\pm}$ are not sensitive to the values of $Q^2$,
for example, in fit II, the value corresponding to $Q^2=1.75 \,
GeV^2$ is about $2\%$ lower than the one at $Q^2=5 \,GeV^2$. The
situation is very different at lower $\varepsilon$ region. The
value of $R^{\pm}$ would reach more than $ 10\%$ at moderate
$Q^2\, \geq \, 1.75 \,GeV^{2}$, for both fits II and III, in the
region $\varepsilon<0.4$. Consequently the new positron
measurements \cite{Arrington04} will very likely shed new light of
the TPE effect in the $ep$ scattering if the experimental
uncertainty can be reduced with $5\%$ and the $\varepsilon$ can
reach 0.4 or below.

In summary, we re-analyze the data of the elastic electron proton
scattering to look for model-independent evidence of the
two-photon-exchange (TPE) effect. In contrast to previous analysis
of \cite{Tg05}, we parametrize the TPE effect  in forms which are
free of kinematical-singularity, in addition to being consistent
with the constraint imposed by crossing symmetry and the charge
conjugation. Moreover, we fix the value of $R=G_E/G_M$ as
determined from the data of the polarization transfer experiment.
We find that the TPE effect in the slope of
$\sigma_R(\varepsilon)$, starting at high $Q^2 \sim 2 \, GeV^2$
values,  is large and comparable with that coming from $G_{E}$ in
the elastic cross section of $ep$ scattering. It becomes dominant
when $Q^2$ reaches $\geq 4 \,GeV^2$. In addition, the TPE effect
is found to behave quasi-linearly in the region of the current
data, i.e., $0.2<\varepsilon<0.95$. It explains why the previous
fits do not find clear evidence for the nonlinearity in the data
of $\sigma_{R}(\varepsilon)$. Hence the fact that the current
elastic $ep$ cross section data shows little nonlinearity with
respect to $\varepsilon$ can not be used to exclude the presence
of the TPE effect.

Our values of $G_{E}$ is much smaller than the results of the
Rosenbluth fit. On the other hand, our values of $G_{M}$ are about
$2\%$ larger than the results of the Rosenbluth fit. The ratio of
the positron-proton and the electron-proton scattering cross section
is also estimated according to our parametrizations. We find that
the value of $R^{\pm}$ would reach more than $ 10\%$ at moderate
$Q^2\, \geq \, 1.75 \,GeV^{2}$ in the region $\varepsilon<0.4$ which
would be of interest for future experimental planning. More precise
data taken outside the region of the current data $\varepsilon<0.2$
or $\varepsilon>0.95$ will be required to
distinguish the different parameterizations of the TPE effect.\\

\noindent{\bf Acknowledgements}\\
\\
This work is partially supported by the National Science Council of
Taiwan under grants nos. NSC095-2112-M022-025 (S.N.Y. and Y.C.C.)
and NSC095-2112-M033-014 (C.W.K.). We thank J. Arrington for
providing us with the data sets and E. Tomasi-Gustafsson for helpful
communications.



{}

\end{document}